\documentclass[mypaper,7pt,twoside]{CoAst}
\usepackage{epsf,graphicx,fancyhdr}
\renewcommand{\chaptermark}[1]{\markboth{#1}{}}

\renewcommand{\headrulewidth}{0pt}
\newcommand{\kopf}{\small\itshape Comm. in Asteroseismology\\ Vol. number, publication date (will be inserted in the production process)}
\newcommand{\Authors}[1]{\begin{center}\normalsize\bf\sf #1 \end{center}}

\renewcommand{\author}[1]{\begin{center}\normalsize\bf\sf #1 \end{center}}
\newcommand{\Address}[1]{\begin{center}\small\sf #1 \end{center}}

\renewenvironment{abstract}{\section*{Abstract}\normalsize\sf}{}
\newcommand{\References}[1]{\begin{flushleft}{\large References\\}\vspace*{2mm}\small #1 \end{flushleft}}

\newcommand{\chapterDSSN}[2]{\chapter[\sf\normalsize #1\\ \footnotesize \hspace*{5mm}by #2 \sf\normalsize][]{#1\\}\rhead[\fancyplain{}{\sf\footnotesize \center{#1}}]{\fancyplain{}{\sffamily\thepage}}\lhead[\fancyplain{\kopf}{\sffamily\thepage}]{\fancyplain{\kopf}{\sf\footnotesize \center{#2}}}}

\newcommand{\figureDSSN}[5]{\begin{figure}[#4]
\centering
\includegraphics*[#5]{#1}
\caption{#2}
\label{#3}
\end{figure}}

\renewcommand{\chaptername}{}
\newcommand{\acknowledgments}[1]{\vspace*{5mm}\noindent\begin{bf}Acknowledgments. \end{bf} #1}

\begin{document}
\sf

\chapterDSSN{Identifying pulsation modes from two-passband
photometry}{Jadwiga Daszy\'nska-Daszkiewicz}
\Authors{Jadwiga Daszy\'nska-Daszkiewicz} \Address{Instytut
Astronomiczny, Uniwersytet Wroc{\l}awski, ul. Kopernika 11, Poland}

\noindent
\begin{abstract}
I discuss a prospect for mode identification from two-passband
photometry of forthcoming BRITE space mission. Examples of
photometric diagnostic diagrams are shown for three types of main
sequence pulsating variables: $\beta$ Cephei, Slowly Pulsating
B-type and $\delta$ Scuti stars. I consider also taking into account
the radial velocity data from simultaneous spectroscopy, which can
be carried out from the ground. With such observations, much better
discrimination of the spherical harmonic degree, $\ell$, can be
accomplished and more constraints on stellar parameters and input
physics can be derived.

\end{abstract}

\noindent
\section{Introduction}

Nowadays space based observations allow detecting oscillation modes
with lower and lower photometric amplitudes. We are already at the
detecton threshold of the order of $10^{-5} - 10^{-6}$ mag,
resulting in a growing number of frequency peaks. There are many
examples of excellent work both observational and theoretical based
on WIRE and MOST data (e.g. Bruntt et al. 2007, Walker et al. 2005,
Barban et al, 2007, Saio et al. 2007). Now we expect similar results
from the just initiated COROT mission.

However, for using these rich frequency data for asteroseismic
modelling, mode identification is a prerequisite. In the case of
main sequence pulsators, we are still far from obtaining very
regular patterns in the oscillation spectra, which could help in
solving the problem. We are also far from nonlinear theory, which
would answer a question about mode selection mechanism. The main
unresolved problem is why most of the theoretically unstable modes
are not observed. Nowakowski (2005) suggested that the dominant
effect of limiting the mode amplitude is a collective saturation of
the opacity driving mechanism, instead of a resonant mode coupling.

In order to identify modes in $\beta$ Cephei, $\delta$ Scuti or
Slowly Pulsating B-type stars, we need additional observables from
multicolour photometry or/and spectroscopy. The photometric method
of mode identification consists in using amplitude ratios and phase
differences in different passbands; it is based on the
semi-analytical formula for the light variation due to linear
pulsation derived by Dziembowski (1977). Balona \& Stobie (1979)
showed that modes with different values of $\ell$ are located in
different regions on the amplitude ratios $vs$ phase differences
diagram. Since then, the method has been applied to various types of
pulsating variables by many authors (Watson 1988, Garrido et al.
1990, Heynderickx et al. 1994). The next important improvement was
including nonadiabatic calculations by Cugier, Dziembowski \&
Pamyatnykh (1994), who applied the method to $\beta$ Cep stars.
Then, Balona \& Evers (1999) emphasized the problem of very high
sensitivity of photometric amplitudes and phases to the treatment of
convection in the case of $\delta$ Sct stars. A photometric
identification of $\ell$ for SPB stars, based on nonadiabatic
calculations, was performed by Townsend (2002). All these works were
done by assuming that the rotation  does not influence pulsation.
However, main sequence pulsators are very often rapid rotators. The
next step in developing the method was its extension to close
frequency modes coupled by a fast rotation by
Daszy\'nska-Daszkiewicz et al. (2002). The photometric method in the
version for long-period g-modes in rotating stars, for which
perturbation approach is no longer adequate, was formulated by
Townsend (2003a) and Daszy\'nska-Daszkiewicz et al. (2007).

A few years ago, Daszy\'nska-Daszkiewicz et al. (2003,2005a)
proposed a new method of the identification of $\ell$, which uses
the amplitudes and phases themselves and combines photometry, and
radial velocity data. The method allows also to extract
simultaneously a new asteroseismic probe, which yields constraints
on stellar parameters and input physics, e.g., convection,
opacities.

BRITE (BRIght Target Explorer) is the first space-based mission
which will perform two-colour photometry of bright stars. It will
give an opportunity of not only detecting low-level oscillations but
also of identifying their degrees $\ell$. However, much more could
be achieved, if ground-based spectroscopic observations are
organized simultaneously. Then another type of information,
contained in the radial velocity and line profile variations, would
be supplied.

The aim of this paper is to show what can be done for mode
identification from observations with the two BRITE passbands. I
will recall the basic formulae and show examples of photometric
diagnostic diagrams for models of $\beta$ Cephei, SPB and $\delta$
Scuti stars. Then I will present diagrams which include amplitudes
and phases from photometric and radial velocity variations. Finally,
I will discuss uncertainties arising from effects of rotation,
convection and atmospheric models. A summary and conclusions are
given in the last section.

\noindent
\section{Mode identification from photometry}

In order to calculate photometric amplitudes and phases, two inputs
are needed. They come from
\begin{itemize}
\item nonadiabtic theory of stellar pulsation,
\item models of stellar atmospheres.
\end{itemize}
We assume linear pulsation theory, which is adequate because of
small mode amplitudes in main sequence pulsators, and we use
temporally static, plane parallel atmosphere, which is justified
because the eigenfunctions of the considered modes are nearly
constant in the atmosphere.

Let us consider a pulsation mode in the zero-rotation approximation,
the geometry of which can be described by a single spherical
harmonic, $Y_\ell^m$, with the degree $\ell$ and the azimuthal order
$m$. The shape of the radial eigenfunctions of the mode are
determined by its radial order $n$. Then the local radial
displacement is given by
$$\frac{\delta r}R=  \varepsilon
Y_\ell^m(\theta,\phi) {\rm e}^{-{\rm i}\omega t}, \eqno(1)$$
where $\varepsilon$ is the intrinsic mode amplitude, $\omega$ is the
angular pulsation frequency, which of course depends on
($n,\ell,m$); other symbols have their usual meanings. The
corresponding changes of the bolometric flux, ${\cal F}_{\rm bol}$,
and the local gravity, $g$, are given by
$$\frac{ \delta {\cal F}_{\rm bol} } { {\cal F}_{\rm bol} }=
\varepsilon f Y_\ell^m(\theta,\phi) {\rm e} ^{-{\rm i} \omega
t},\eqno(2)$$
and
$$\frac{\delta g_{\rm eff}}{g_{\rm eff}} = - \varepsilon \left( 2 + \frac{\omega^2 R^3}{G M}
\right) Y_\ell^m(\theta,\phi) {\rm e} ^{-{\rm i} \omega t}.
\eqno(3)$$
The complex parameter, $f$, describes the ratio of the local flux
perturbation to the radial displacement at the level of the
photosphere and it is obtained from nonadiabatic calculations.

The complex photometric amplitudes of the light variation in the
passband $\lambda$ due to a pulsation mode with frequency $\omega$
can be written in the following form
$${\cal A}_{\lambda}(i) = -1.086 \varepsilon Y_{\ell}^m(i,0) b_{\ell}^{\lambda}
(D_{1,\ell}^{\lambda}+D_{2,\ell}+D_{3,\ell}^{\lambda})\eqno(4)$$
where
$$D_{1,\ell}^{\lambda} = \frac14  f \frac{\partial \log ( {\cal
F}_\lambda |b_{\ell}^{\lambda}| ) } {\partial\log T_{\rm{eff}}} ,
\eqno(5a)$$
$$D_{2,\ell} = (2+\ell )(1-\ell ), \eqno(5b)$$
$$D_{3,\ell}^{\lambda}= -\left( 2+ \frac{\omega^2 R^3}{GM} \right)
 \frac{\partial \log ( {\cal F}_\lambda |b_{\ell}^{\lambda}|
) }{\partial\log g_{\rm eff}^0} \eqno(5c)$$
and $i$ is the inclination angle. The partial derivatives of $\log (
{\cal F}_\lambda |b_{\ell}^{\lambda}|)$ over effective temperature
and gravity are derived from atmospheric models and
$$b_{\ell}^\lambda=\int_0^1 h_\lambda(\mu) \mu P_{\ell}(\mu)
d\mu, \eqno(6)$$
is the disc-averaging factor, containing the information about the
visibility of the mode with a given degree $\ell$. The integrals
$b_\ell^\lambda$ are weighted by the limb-darkening law,
$h_\lambda(\mu)$. The term $D_{1,\ell}^\lambda$ describes the
temperature effects, the term $D_{2,\ell}$ stands for the
geometrical effects, and the influence of gravity changes is
contained in the term $D_{3,\ell}^\lambda$. The terms
$D_{1,\ell}^\lambda$ and $D_{3,\ell}^\lambda$ include the
perturbation of the limb-darkening, and their $\ell$-dependence
arises from the nonlinearity of the limb-darkening law.
\figureDSSN{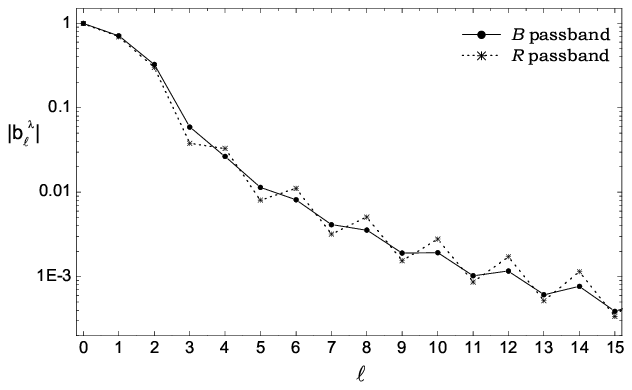}{ The disc-averaging factor, $b_\ell^\lambda$,
as a function of the spherical harmonic degree, $\ell$, for the $BR$
Johnson filters. The adopted stellar parameters are $\log T_{\rm
eff}=3.875$ and $\log g=4.0$.}
{fig1}{!ht}{clip,height=5.5cm,width=\textwidth}

The amplitudes and phases of the light variation are given by
$|{\cal A}_{\lambda}|$ and ${\rm arg}({\cal A}_{\lambda})$,
respectively. Having these numbers, we can construct photometric
diagnostic diagrams in the form  $A_x/A_y~ vs.~ \varphi_x -
\varphi_y$, where $x$ and $y$ denote passbands. These observables
are independent of the intrinsic amplitude, $\varepsilon$,
inclination angle, $i$, and azimuthal order, $m$, because the
product $\varepsilon Y_{\ell}^m$ drops out. These is an advantage
for the identification of $\ell$ but also a disadvantage because the
order $m$ is beyond of the reach of the photometric method.

In this section I will give examples of photometric diagrams in the
$BR$ Johnson filters which are not very different from the BRITE
passbands: BT1 (390-460 nm) and BT2 (550-700 nm). The well known
property of the factor $b_{\ell}^{\lambda}$ is that it decreases
very rapidly with growing degree $\ell$; this can be seen from Fig.
1. In this paper I will consider modes with degrees up to $\ell=6$.
All calculations were done with the Warsaw-New Jersey evolutionary
code and nonadiabatic pulsation code of Dziembowski (1977). I used
OPAL opacity tables of Iglesias \& Rogers (1996) and the solar
chemical composition of Grevesse \& Noels (1993), assuming the
metallicity $Z=0.02$. I adopted Kurucz atmospheric models in the
NOVER-ODFNEW version (Castelli \& Kurucz 2004), which have more
smooth flux derivatives than the standard Kurucz models, and the
Claret nonlinear formula for the limb-darkening law. The standard
atmospheric metallicity, [m/H]=0.0, and the microturbulence
velocity, $\xi_t=2$ km/s, were assumed.

\subsection{$\beta$ Cephei pulsators}

\figureDSSN{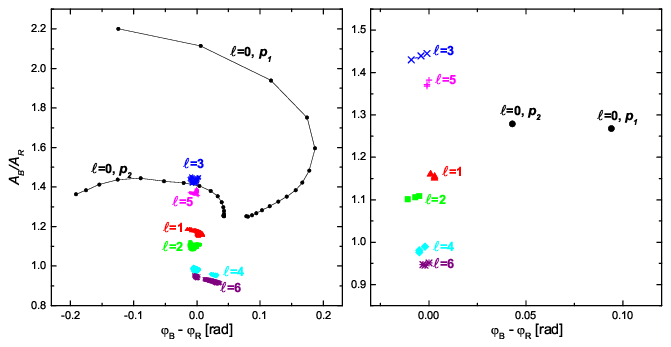}{The locations of unstable modes with degrees
$\ell$ up to 6 for $\beta$ Cephei star models of 12 $M_\odot$ on the
diagnostic diagrams involving Johnson $B$ and $R$ filters. The left
panel contains all models in the main sequence evolutionary phase,
and the right panel, the model with $\log T_{\rm eff}=4.400$ and
$\log g=3.89$.} {fig2}{!ht}{clip,width=\textwidth}
I considered stellar models with a mass of $M=12 M_\odot$ during
main sequence phase of evolution, corresponding to the temperature
range of $\log T_{\rm eff}=4.445 - 4.347$. In Fig. 2, the
photometric diagram in the $BR$ passband is presented. The left
panel shows all modes which become unstable between ZAMS and TAMS
and the right panel shows unstable modes for only one stellar model
with $\log T_{\rm eff}=4.400$ and $\log g=3.89$. Modes with
different degrees $\ell$ are located in separated regions. The
radial modes are spread over a wide range of the amplitude ratios
and phase differences, whereas the nonradial modes are concentrated
in small areas. This behaviour results from different contributions
of the temperature ($D_{1,\ell}^\lambda f$) and the geometrical
effects ($D_{2,\ell}$) to the light variation
(Daszy\'nska-Daszkiewicz et al. 2002)

\subsection{Slowly Pulsating B-type stars}

In order to calculate SPB oscillation, I chose the models with $M=5
M_\odot$ on the main sequence, which include $\log T_{\rm eff}$ from
4.235 to 4.134. Positions of modes with different values of $\ell$
on the photometric $BR$ diagram are shown in Fig. 3. Again, the left
panel contains all unstable modes, and the right panel, one model
with $\log T_{\rm eff} =4.195$ and $\log g=4.02$. The well-known
property of this type of diagrams is the zero phase difference and
the same amplitude ratio for all modes with $\ell=1$. This is
because the light variation of the $\ell=1$ mode in SPB models is
totaly dominated by the temperature effects. We can see an
overlapping of domains with different values of $\ell$. This can be
partly removed by considering only one model (the right panel).
\figureDSSN{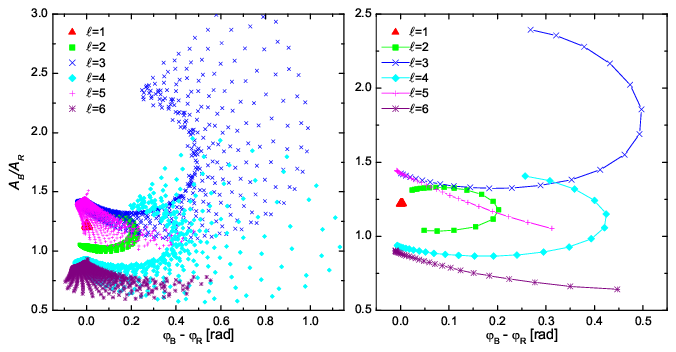}{The same as in Fig. 2 but for Slowly Pulsating
B-type star models of 5 $M_\odot$. The left panel contains all
models in the main sequence evolutionary phase and the right panel,
the model with $\log T_{\rm eff}=4.195$ and $\log g=4.02$.}
{fig3}{!ht}{clip,width=\textwidth}
\figureDSSN{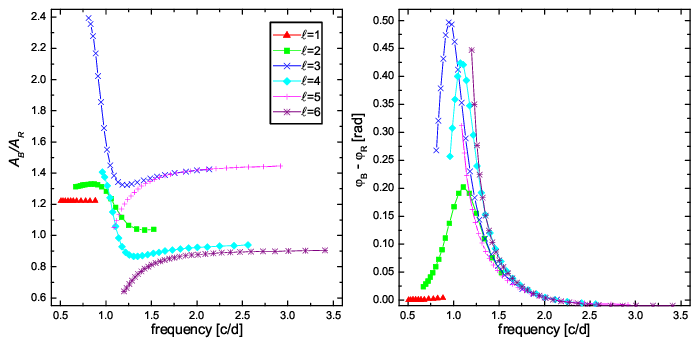}{Amplitude ratios (on the left) and phase
differences (on the right) in the $BR$ bands as a function of
oscillation frequency for the SPB model with $\log T_{\rm
eff}=4.195$ and $\log g=4.02$.} {fig4}{!ht}{clip,width=\textwidth}
Another instructive information can be drawn from instability
conditions. Fig.4 shows the amplitude ratio and the phase difference
as a function of frequency for the model with $\log T_{\rm eff}
=4.195$ and $\log g=4.02$. As one can see, modes with $\ell=1$ are
unstable only for the lowest frequencies ($\nu<0.9$ c/d). The
instability is shifted to the higher frequencies for higher
$\ell$-modes, e.g., the $\ell=5$ modes become unstable for $\nu>1.1$
c/d.

\subsection{$\delta$ Scuti pulsators}

As representatives of $\delta$ Sct pulsator, I took models with $M=2
M_\odot$ in the main sequence phase. The corresponding $\log T_{\rm
eff}$ range is (3.963, 3.854). All calculations were made under
assumptions of the mixing length theory and the convective flux
freezing approximation. The mixing length parameter was $\alpha_{\rm
conv}=0.0$, i.e., an assumption of inefficient convective transport.
The photometric $BR$ diagram for this type of pulsators is presented
in Fig. 5. All unstable modes of $M=2 M_\odot$ models are shown on
the left hand side, and modes in the ($\log T_{\rm eff}=3.909,~\log
g=4.02$) model on the right hand side. As we can see, there is some
overlap, especially for the $\ell\le 2$ modes. Fixing the model
parameters helps in removing this ambiguity.
\figureDSSN{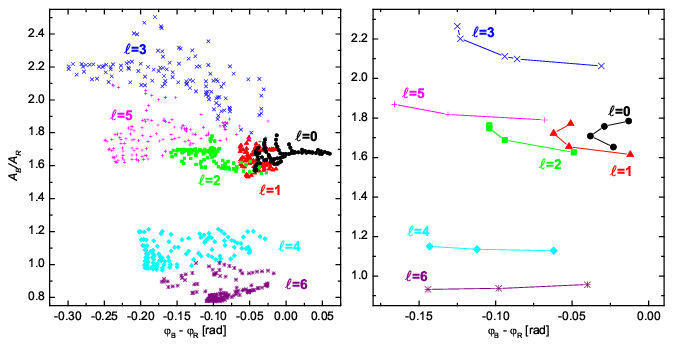}{The same as in Fig. 2 but for $\delta$ Scuti
star models of 2 $M_\odot$. The left panel containes all models in
the main sequence evolutionary phase, and the right panel, the model
with $\log T_{\rm eff}=3.909$ and $\log g=4.02$.}
{fig5}{!ht}{clip,width=\textwidth}
\figureDSSN{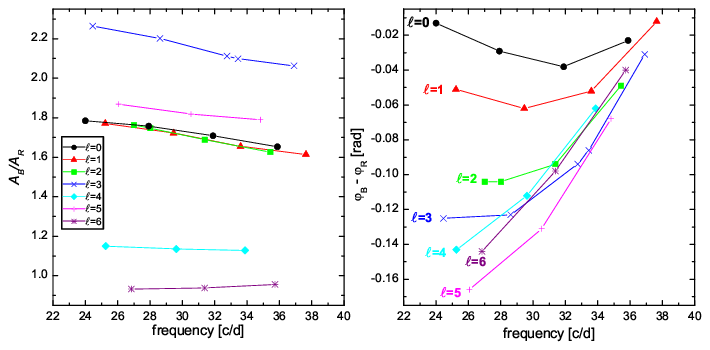}{Amplitude ratios (on the left) and phase
differences (on the right) in the $BR$ bands as a function of
oscillation frequency for the $\delta$ Sct model with $\log T_{\rm
eff}=3.909$ and $\log g=4.02$.} {fig6}{!ht}{clip,width=\textwidth}

The common opinion about $\delta$ Sct pulsation modes is that the
main information about the $\ell$ values is contained in the phase
differences. In fact, this is true only for low degree modes with
$\ell \le 2$, but for higher $\ell$'s we can learn much also from
the amplitude ratios. In Fig. 6, the amplitude ratio (the left
panel) and phase difference (the right panel) as a function of
oscillation frequency are depicted. As we can see, not much can be
we achieved for low degree modes also from the $A_B/A_B~vs.$
frequency plot because the $\ell\le 2$ modes are excited with very
close frequencies. Another interesting property to note is that for
very high frequency mode ($\nu>35$ c/d), the phase differences are
almost the same for all degrees $\ell$.

\noindent
\section{Adding radial velocity measurements}

The radial velocity variation averaged over stellar disc is
expressed by the well-known Dziembowski's (1977) formula
$$V_{rad}(i)= {\rm i}\omega R \varepsilon Y_\ell^m(i,0)
\left( u_{\ell}^{\lambda} + \frac{GM}{R^3\omega^2}v_{\ell}^{\lambda}
\right), \eqno(7)$$
where
$$u_{\ell}^{\lambda} = \int_0^1 h_\lambda(\mu) \mu^2 P_{\ell}(\mu)
d\mu, \eqno(8a)$$
and
$$v_{\ell}^{\lambda} = \ell \int_0^1 h_\lambda(\mu) \mu \left(
P_{\ell-1}(\mu) - \mu P_{\ell}(\mu) \right)d\mu. \eqno(8b)$$
From observations, the radial velocity variations are determined by
calculating the first moment, ${\cal M}_1^{\lambda}$, of a well
isolated spectral line.

In this section, I show examples of diagnostic diagrams constructed
from the radial velocity variation and the light variation in the
Johnson $R$ filter. Unstable oscillation modes for $\beta$ Cep, SPB
and $\delta$ Sct star models considered in the previous section are
presented in Fig 7, 8 and 9, respectively.
\figureDSSN{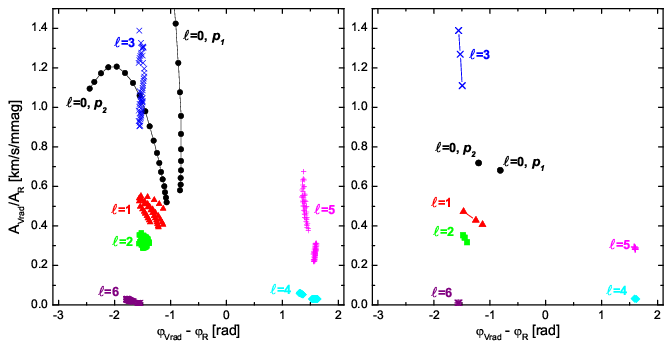}{The location of unstable modes with degrees
$\ell$ up to 6 for $\beta$ Cep models of 12 $M_\odot$ on the
diagnostic diagrams constructed with amplitudes and phases for the
$R$ passband and the radial velocity variations. As in Fig. 2, the
left panel shows all main sequence models, and the right panel, one
model with $\log T_{\rm eff}=4.400$ and $\log g=3.89$.}
{fig7}{!ht}{clip,width=\textwidth}
\figureDSSN{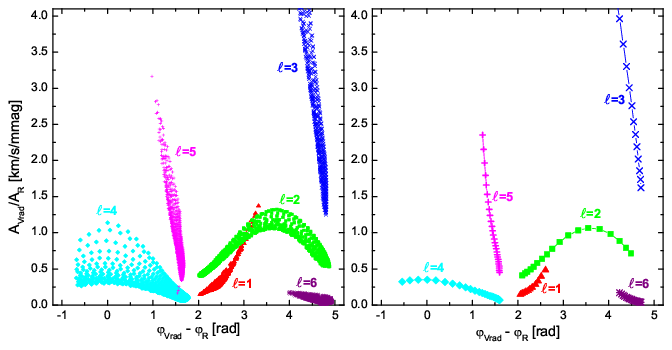}{The same as in Fig. 7 but for SPB models of 5
$M_\odot$. In the right panel unstable modes for the model with
$\log T_{\rm eff}=4.195$ and $\log g=4.02$ are plotted.}
{fig8}{!ht}{clip,width=\textwidth}
\figureDSSN{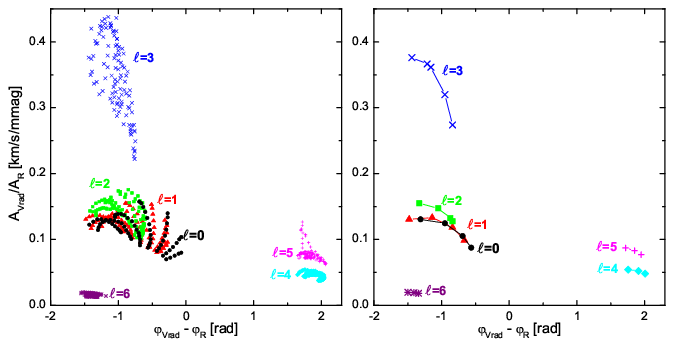}{The same as in Fig. 7 but for $\delta$ Sct
models of 2 $M_\odot$. In the right panel unstable modes for the
model with $\log T_{\rm eff}=3.909$ and $\log g=4.02$ are plotted.}
{fig9}{!ht}{clip,width=\textwidth}
As we can see, the configuration of  $\ell$ modes domains differs
from that in the photometric diagrams. In particular, there are
larger phase differences between light and radial velocity variation
than between two passnands.

Moreover, with two photometric passbands and the radial velocity
data, one can apply the method of simultaneous extracting from
observations the degree $\ell$ and the nonadiabatic parameter $f$.
Determination of the empirical value of $f$ allows to avoid, in the
process of the $\ell$-identification, the input from pulsation
models, which still needs many improvements, for example the
pulsation-convection interaction, opacity tables, effects of
diffusion, mixing etc. On the other hand, the parameter $f$
constitutes a new asteroseismic probe, giving information on
subphotospheric layers and is complementary to oscillation
frequencies determined by stellar interior. Comparing empirical and
theoretical values of $f$, one can draw conclusions about the
efficiency of convection in $\delta$ Sct stars
(Daszy\'nska-Daszkiewicz et al 2003,2005b) or about opacities in
$\beta$ Cep stars (Daszy\'nska-Daszkiewicz et al. 2005a).

\noindent
\section{Uncertainties}

There are many sources of uncertainties in all results presented in
this paper. Firstly, all calculation were done assuming the zero
rotation approximation. Effects of rotation can spoil the nice mode
separation in the diagnostic diagrams, making them dependent on the
inclination, rotation rate and the azimuthal order $m$. One such
effect is rotational mode coupling. It takes place if the frequency
distance between modes, with the degrees $\ell$ differing by 2 and
the same $m$, is of the order of the rotation frequency. In such a
situation, the photometric amplitude for the coupled mode has to be
calculated as superposition of all mode amplitudes satisfying the
conditions $\ell_k=\ell_j+2$ and $m_k=m_j$. The effect of rotational
mode coupling on the photometric diagrams for $\beta$ Cep star was
studied by Daszy\'nska-Daszkiewicz et al. (2002). Examples for
$\delta$ Sct stars can be found in Daszy\'nska-Daszkiewicz (2007).
Another case when the rotation has to be included is when the
pulsational frequency, $\omega$, is of the order of the rotational
frequency, $\Omega$, so that the perturbation approach fails. It
happens often in the case of the rapidly rotating SPB stars, in
which high order gravity modes are excited. Such slow modes are no
longer described by the spherical harmonics, and more complicated
formalisms are needed. One possibility is the use of the traditional
approximation which allows expressing the angular dependence of
eigenfunctions by the Hough functions (e.g. Lee \& Saio 1997,
Townsend 2003b). The formula for the light variation due to low
frequency oscillation was given by Townsend (2003a).
Daszy\'nska-Daszkiewicz et al. (2007) discussed a prospect for mode
identification from diagnostic diagrams and derived the expression
for the radial velocity variation.

Another uncertainty comes from the input physics and atmospheric
models. The effect of metallicity parameter, $Z$, and opacities on
the diagnostic diagrams for $\beta$ Cep star models was considered
by Cugier et al. (1994). For a higher value of $Z$, the separation
of modes with different $\ell$'s is much better. Similarly,
computations with the OP tables, instead of OPAL, lead to a little
better $\ell$ discrimination, especially the radial modes are much
more spread. Then, Cugier \& Daszy\'nska (2001) checked the effect
of the atmospheric metallicity parameter, [m/H], and the
microturbulence velocity, $\xi_t$, on the diagnostic properties of
photometric diagrams. For $\beta$ Cep stars these parameters have
negligible impact.

The main problem in applying the method of diagnostic diagrams to
$\delta$ Sct modes is that photometric amplitudes and phases exhibit
a strong dependence on the treatment of convection (Balona \& Evers
1999). To circumvent this problem, Daszy\'nska-Daszkiewicz et al.
(2003, 2005b) invented the method of simultaneous determination of
degree $\ell$ and the nonadiabatic parameter $f$ from observations.
Then we can identify the $\ell$-value independently of the pulsation
models and, by comparing empirical and theoretical $f$ values,
constraints on convection can be inferred. The result was that the
convective transport in $\delta$ Sct stars studied by us is rather
inefficient. The progress in modelling $\delta$ Sct pulsation with
time-dependent convection treatment was achieved e.g. Grigah\`cene
et al. (2005), Dupret et al. (2005a) and Dupret et al. (2005b).

In the case of $\delta$ Sct models, the uncertainties in atmospheric
models can play much more important role than in the B-type
pulsators. In Kurucz standard models, the flux derivatives over
effective temperature and gravity are not smooth at the temperature
where convective transport becomes important. This is because of
using an overshooting approximation that moves the flux higher in
the atmosphere, above the top of the nominal convection zone. In the
NOVER-ODFNEW models computed by F. Castelli the problem of
non-smooth derivatives does not exist. There are also NEMO models
({\b Ne}w {\bf Mo}del Grid of Stellar Atmospheres) which include
different treatment of convection. The models were computed by the
Vienna group with modified versions of the Kurucz ATLAS9 code. The
grids have smaller steps in $T_{\rm eff}$ and $\log g$ than in
Kurucz's computations, and the flux derivatives are perfectly
smooth. The effect of using various atmospheric models in the
calculation of $\delta$ Sct observables was discussed by
Daszy\'nska-Daszkiewicz et al. (2004) and Daszy\'nska-Daszkiewicz
(2007).

\noindent
\section{Conclusions}

There is a potential for mode identification from the BRITE
photometry. As we could see, two-colour information can yield some
constraints on the spherical harmonic degree, $\ell$, of the
oscillation modes excited in main sequence pulsators.

However, space observations should be followed up by simultaneous
ground-based spectroscopy. The advantages of adding the
spectroscopic variations is obvious. By combing photometry and the
radial velocity data, we will improve significantly the
discrimination of $\ell$ and infer better seismic constraints on
stellar parameters and input physics. Moreover, from the line
profile variations the identification of the azimuthal order, $m$,
becomes possible. With such unprecedented data, we can hope for a
great step in asteroseismology of main sequence pulsators with the
BRITE-Constellation.

 \acknowledgments{
The author thanks Werner Weiss for inviting her to participate in
the BRITE Workshop and Miko{\l}aj Jerzykiewicz for carefully reading
the manuscript. This work was supported by the Polish MNiSW grant
No. 1 P03D 021 28.}

\References{
Balona L. A., Evers E. A., 1999, MNRAS 302, 349\\
Balona L. A.,Stobie R. S., 1979, MNRAS 189, 649\\
Barban C., Matthews J. M., De Ridder j., 2007, A\&A 468, 1033\\
Bruntt H., Su\'arez J. C., Bedding T. R., et al., 2007, A\&A 461, 619\\
Castelli F., Kurucz R. L., 2004,  the Proceedings of the IAU Symp. No 210,
in Modelling of Stellar Atmospheres, eds. N. Piskunov et al. 2003, poster A20\\
Cugier H., Daszynska J., 2001, A\&A 377, 113\\
Cugier H., Dziembowski W. A., Pamyantykh A. A., 1994, A\&A 291, 143\\
Daszy\'nska-Daszkiewicz J., Dziembowski W. A., Pamyantykh A. A., Goupil M-J., 2002, A\&A 392, 151\\
Daszy\'nska-Daszkiewicz J., Dziembowski W. A., Pamyatnykh A. A., 2003, A\&A  407, 999\\
Daszy\'nska-Daszkiewicz J., Dziembowski W. A., Pamyatnykh A. A.,
Breger M., Zima W., 2004, in The A-Star Puzzle, IAU Symposium, No.
224, p.853-859, eds. J. Zverko, J. Ziznovsky, S.J. Adelman, and W.W.
Weiss, Cambridge University Press \\
Daszy\'nska-Daszkiewicz J., Dziembowski W. A., Pamyatnykh A. A., 2005a, A\&A  441, 641\\
Daszy\'nska-Daszkiewicz J., Dziembowski W. A., Pamyatnykh A. A., Breger M., Zima W., Houdek G., 2005b, A\&A 438, 653\\
Daszy\'nska-Daszkiewicz J., 2007, Comm. Asteroseis. 150, 32\\
Daszy\'nska-Daszkiewicz J., Dziembowski W. A., Pamyatnykh A. A., 2007, Acta Astronomica  57, 11\\
Dziembowski W. A., 1977, Acta Astron. 27, 203\\
Dupret M.-A., Grigahc\`ene A., Garrido, Gabriel M., Scuflaire R., 2005a, A\&A 435, 927\\
Dupret M.-A., Grigahc\`ene A., Garrido R., De Ridder J., Scuflaire
R., Gabriel M., 2005b, MNRAS 361, 476\\
Garrido R., Garcia-Lobo E., Rodriguez, E., 1990, A\&A 234, 262\\
Grigahc\`ene A., Dupret M.-A., Gabriel M., Garrido R., Scuflaire R., 2005, A\&A 434, 1055\\
Heynderickx D., Waelkens C., Smeyers P., 1994, A\&AS 105,447\\
Iglesias, C. A., Rogers, F. J., 1996, ApJ 464, 943\\
Nowakowski, R., 2005, Acta Astron. 55, 1\\
Saio H., Cameron C., Kusching R., et al., 2007, Comm. Asteroseis. 150, 215\\
Townsend R. H. D., 2002, MNRAS 330, 855\\
Townsend R. H. D., 2003a, MNRAS 343, 125\\
Townsend R. H. D., 2003b, MNRAS, 340, 1020\\
Walker G. A. H., Kuschnig R., Matthews J. M., et al., 2005, ApJ, 635, L77\\
Watson R.D., 1988, Ap\&SS 140, 255 }

\end{document}